\documentclass[journal=jpccck,manuscript=article,layout=twocolumn]{achemso}
\usepackage[toc]{appendix}
\usepackage{amsmath}
\usepackage{amssymb}
\usepackage{graphicx}
\usepackage{dcolumn}
\usepackage{bm}
\usepackage{titlesec}
\usepackage{natbib}
\usepackage{color}

\titleformat{\paragraph}
{\normalfont\normalsize\bfseries}{\theparagraph}{1em}{}
\titlespacing*{\paragraph}
{0pt}{3.25ex plus 1ex minus .2ex}{1.5ex plus .2ex}

\title{ An Almost Analytical Approach to Simulating 2D Electronic Spectra}

\author{Pallavi Bhattacharyya}
 \affiliation{Department of Chemistry and Chemical Biology, Cornell University, Ithaca, New York 14853, USA}
\author{Nandini Ananth}%
\affiliation{Department of Chemistry and Chemical Biology, Cornell University, Ithaca, New York 14853, USA
}%
\email{ananth@cornell.edu.}

\date{\today}

\begin{document}
\maketitle
\twocolumn[
\begin{@twocolumnfalse}

\begin{abstract}

We introduce an almost analytical
method to simulate 2D electronic spectra
as a double Fourier transform of the 
the non-linear response function (NRF) 
corresponding to a particular
optical pulse sequence.
We employ a unitary transformation to represent the 
total system Hamiltonian in a stationary 
basis that allows us to separate contributions from 
decoherence and phonon-mediated population relaxation 
to the NRF.
Previously, one of us demonstrated the use 
of an analytic, cumulant expansion approach 
to calculate the decoherence term. 
Here, we extend this idea to obtain
an accurate expression for the population 
relaxation term, a significant 
improvement over standard quantum 
master equation-based approximations.
We numerically demonstrate the accuracy of 
our method by computing the photon echo spectrum 
of a two-level system coupled to a thermal bath,
and we highlight the mechanistic insights obtained
from our simulation.
\end{abstract}
  \end{@twocolumnfalse}
]
\section{Introduction}
2D Electronic Spectroscopy is a four wave 
mixing technique \cite{mukamelbook, Zhang1998, Hybl1998307, 
Hybl2001, Stiopkin2004, Tonu2006, chojccs, Read04092007, 
Cheng2007, ginsberg,   Scholes2014, Jakub2016}
that can, uniquely, report on excitonic transitions, 
couplings and relaxation pathways. 
These measurements provide invaluable 
insights into the mechanisms of energy transport 
in biological processes including, most notably, 
photosynthetic light harvesting complexes.
However, theoretical simulations are essential to correctly 
interpret measured 2D electronic spectra by disentangling the 
contributions from different mechanistic 
pathways~\cite{jpcbcho,Zigmantas22082006, pisliakov1, yccheng,chenshiq}.

Existing theoretical approaches can be broadly classified
into two categories. 
The first includes methods developed to calculate net 
nonlinear polarization using either nonperturbative 
approaches~\cite{pisliakov1, pisliakov2, geva2006a} or time-nonlocal
quantum master equation based approaches~\cite{yccheng, Domcke}.
However, nonperturbative methods provide limited molecular insights 
and in general, the calculated nonlinear polarization must be subjected 
to significant post-processing to extract the signal due to a 
particular pulse sequence~\cite{Seidner1995}.
The second category of methods directly calculate 
nonlinear response functions (NRFs) that correspond to 
a particular pulse sequence. Existing approaches include 
analytic perturbative methods based on early 
work~\cite{mukamel1993, Cho2001},
however they rely on {\it adhoc} approximations 
including an artificial separation of decoherence 
and population relaxation contributions to the NRF 
making them inaccurate descriptors of dynamics 
particularly at short times \cite{jpcbcho}. 
Liouville space heirarchical equations of motion 
have also been used to calculate NRFs 
numerically~\cite{chenshiq}, however, 
these methods are computationally expensive
and scale poorly with system dimensionality.

In this paper, we introduce a novel, near analytic,
computationally efficient method for the theoretical
calculation of NRFs. 
We focus on the simulation of a 
2D Photon Echo spectrum~\cite{mukamelarpc, SchlauCohen20111, 
Lee1462, Zigmantas22082006} but our approach 
is general and can be trivially extended to 
other types of measurements.
Our method is derived through a series of well defined 
approximations and has several key features: 
a) First, we use a unitary transformation, introduced 
previously, \cite{pallavi1,pallavi2} to map adiabatic states 
to a stationary basis, that allows us to rigorously 
decouple decoherence from the exciton relaxation dynamics,
b) Second, we treat dynamics during the coherence and 
rephasing times ($t_1$ and $t_3$, respectively), and the 
population time $(t_2)$ with the same level of approximation
making our approach accurate at both short and long times,
c) Third, the decoherence contribution to the NRF is evaluated 
analytically for systems where the bath is well described 
by either an Ohmic or a Debye spectral distribution. The 
population relaxation term is evaluated through a series
of simple numerical integrations.
d) Finally, we treat doubly excited states populated 
in the Excited State Absorption (ESA) pathway~\cite{jpcbcho} 
on an even footing with singly excited states with no 
additional approximations.
Taken together, these features render this approach 
very powerful and the near analytic formulation makes it 
computationally inexpensive and easy to implement.
By properly separating contributions to the spectrum 
from decoherence and population relaxation pathways,
we are able to provide necessary mechanistic insights.

The paper is organized as follows. 
First, in Section~\ref{Hamiltoniansection}, 
we introduce the stationary basis and the unitary
mapping transformation employed 
to represent the total Hamiltonian 
for an $n$-level system coupled to a 
thermal bath in this framework.
Next, in Section~\ref{spectroscopy},
we briefly review NRF and the computation
of 2D photon echo electronic spectrum.
We then introduce our approach for the 
Stimulated Emission (SE) pathway in 
Section~\ref{SE} outlining the calculation
of both the decoherence contribution and 
our new approach to evaluate the population 
relaxation contribution.
In Section~\ref{othertwo}, we provide similar 
outlines for both the Ground State Bleaching 
(GSB) and Excited State Absorption (ESA) pathways
both of which contribute to the 2D photon 
echo spectrum.
We then demonstrate the results of our simulation
for a model two-level system and discuss the key
insights obtained in Section~\ref{rdsection}.

\section{Stationary Basis} 
\label{Hamiltoniansection}
The quantum mechanical Hamiltonian for an $n$-level 
system where each state is linearly coupled to a thermal bath 
of harmonic oscillators can be written as
\begin{equation}
\bar{H}=\epsilon_{g}| g \rangle \langle g |+
H_\text{ad}(\textbf{Q}) + H_\text{ph},
\label{Hamiltonian}
\end{equation}
where
\begin{eqnarray}
\nonumber
H_\text{ad}(\textbf{Q})&=& \sum_{j} \epsilon_{j} | j \rangle \langle j | + 
\sum_{i,j;i\neq j}J_{ij} |i \rangle \langle j | \\ 
&+&\sum_{j} Q_{j} |j \rangle \langle j|, 
\label{adia_ham}
\end{eqnarray}
and 
\begin{equation}
H_{ph}=\sum_{j,b}\frac{1}{2}\left(\frac{p_{jb}^{2}}{m_{jb}}+
m_{jb} \omega_{jb}^{2} q_{jb}^{2}\right).
\label{phonon_ham}
\end{equation}

In Eq.~\ref{adia_ham}, $i$ and $j$ label the 
local first excited states, $\epsilon_{j}$ is the 
energy of the $j^{th}$ state, $J_{ij}$ is the 
electronic coupling between the $i^{th}$ and $j^{th}$ states,
and $\epsilon_{g}$ is the ground state 
energy of the full system.
Further, in Eq.~\ref{adia_ham} and Eq.~\ref{phonon_ham}, 
$Q_{j}=\sum_{b}m_{jb} \nu_{jb}q_{jb}$, where 
$m_{jb}$, $q_{jb}$, $p_{jb}$, and $\omega_{jb}$ 
are, respectively, the mass, position, momentum, 
and angular frequency associated with the $b^{th}$ 
harmonic bath mode coupled to the $j^{th}$ state 
of the system. 

We now define a set of adiabatic 
eigenfunctions~\cite{pallavi1,pallavi2}
such that 
\begin{equation}
H_{ad}(\textbf{Q})|m(\textbf{Q})\rangle
=\varepsilon_{m}(\textbf{Q})|m(\textbf{Q})\rangle, 
\label{adiabaticbasis}
\end{equation}
where we introduce the notation $\mathbf{Q}=\{Q_{j}\}$.
We recognize that the $\mathbf{Q}$-dependent adiabatic 
eigenfunctions do not commute with momentum operators 
in $H_{ph}$.  Therefore, we introduce a new stationary basis,
$|m(\textbf{Q}=\textbf{0})\rangle$, that is $\mathbf{Q}$-independent
and that we will denote simply as $|m\rangle$ in the remainder of this 
manuscript. 
We then define a unitary transformation from the adiabatic basis
to our stationary basis~\cite{pallavi1,pallavi2},
\begin{equation}
|m (\textbf{Q}) \rangle =U(\textbf{Q})|m \rangle.
\label{utransf}
\end{equation}

Defining $H=U^{\dag}(\textbf{Q})\bar{H}U(\textbf{Q})$, 
the unitary transformation of the Hamiltonian 
in Eq.~\ref{Hamiltonian}, and introducing two 
physically reasonable approximations, we obtain 
\begin{equation}
H=H_0+H_\text{na}, 
\label{ut_ham}
\end{equation}
where
\begin{eqnarray}
H_{0}=\epsilon_{g}| g \rangle \langle g | +
\sum_{m}\varepsilon_{m}(\textbf{Q})|m\rangle \langle m | 
+ H_\text{ph}
\label{H0}
\end{eqnarray}
is a diagonal matrix in the stationary state basis
and $H_\text{ph}$ is defined previously in 
Eq.~\ref{phonon_ham}.
The part of the Hamiltonian that drives 
nonadiabatic transitions in Eq.~\ref{ut_ham} is 
defined as 
\begin{equation}
H_\text{na}=\frac{1}{2}\sum_{j}(P_{j} A^{j}(\textbf{0})
+A^{j}(\textbf{0})P_{j}),
\label{Hna}
\end{equation}
where $P_{j}=\sum_{b} \nu_{jb}p_{jb}$
and the matrix elements of the 
nonadiabatic coupling vector are 
defined as 
\begin{equation}
A^{j}_{n,m}(\textbf{0})=
-i  \frac{\langle n (\textbf{0})|j \rangle 
\langle  j | m(\textbf{0}) \rangle}{\varepsilon_{n}
(\textbf{0})-\varepsilon_{m}(\textbf{0})}.
\end{equation}

The two approximations mentioned above are 
both used to derive the nonadiabatic Hamiltonian 
in Eq.~\ref{Hna}.
Applying an exact unitary transformation to the 
Hamiltonian in Eq.~\ref{Hamiltonian}, we obtain 
a nonadiabatic coupling vector where 
the $j^\text{th}$ component is defined as 
\begin{equation}
A^{j}(\textbf{Q})=\sum_{n,m}A^{j}_{n,m}(\textbf{Q})
|n\rangle \langle m |,
\label{Aoperator}
\end{equation}
with matrix elements,
\begin{eqnarray}
\nonumber
A^{j}_{n,m}(\textbf{Q})&=& -i  \langle n (\textbf{Q})|
\frac{\partial}{\partial Q_{j}} m(\textbf{Q}) \rangle\\
&=& -i  \frac{\langle n (\textbf{Q})|j \rangle 
\langle  j | m(\textbf{Q}) \rangle}
{\varepsilon_{n}(\textbf{Q})-\varepsilon_{m}(\textbf{Q})},
\label{Aj1}
\end{eqnarray}
where we use the Hellmann-Feynman theorem to obtain the 
second equality.
Since both the numerator, involving overlaps 
between stationary states and local excited states,
and the denominator, the energy gap term, are likely to be robust with respect to phonon-induced fluctuations,
we first approximate the nonadiabatic coupling
vector by its value at $\mathbf{Q}=\textbf{0}$, 
i.e. $A^j(\mathbf{Q})\approx A^j(\textbf{0})$.
Second, treating the nonadiabatic 
coupling term perturbatively, we assume 
that terms which are second order in $A^j(\mathbf{Q})$ 
are negligible.

\section{Simulating the Photon Echo Spectrum} \label{spectroscopy}

Stimulated photon echo electronic spectroscopy 
is a three pulse UV-vis experiment, with 
phase matching direction 
$\textbf{k}_{S}=-\textbf{k}_{1}+\textbf{k}_{2}+\textbf{k}_{3}$. 
Here, we provide a concise definition for the 
2D Photon Echo spectrum in terms of the 
relevant response functions. 
A more detailed description is available 
in the literature~\cite{jpcbcho,SchlauCohen20111,yccheng}.

We calculate the 2D photon echo 
spectrum from the expression~\cite{SchlauCohen20111}
\begin{align}
\tilde{S}_\text{PE}(\omega_{1},t_{2},\omega_{3}) & =
\int_{0}^{\infty}dt_{3} e^{i \omega_{3} t_{3}}
\int_{-\infty}^{\infty}dt_{1} e^{-i \omega_{1} t_{1}} \nonumber \\ 
& \times  S_\text{PE}(t_{3},t_{2},t_{1}), 
\label{echofourier}
\end{align}
where $\omega_{1}$ and $\omega_{3}$ are 
fourier transform frequencies. 
The time-domain photon echo signal 
in Eq.~\ref{echofourier} is defined 
in terms of response functions,
\begin{align}
S_{PE}(t_{3},t_{2},t_{1}) & =K [R_\text{SE}(t_{3},t_{2},t_{1}) \nonumber \\
& + R_\text{GSB}(t_{3},t_{2},t_{1})
- R^{*}_\text{ESA}(t_{3},t_{2},t_{1})], 
\label{spe}
\end{align}
where $K$ is a common prefactor 
containing the dot products of 
the transition dipole moment unit 
vectors with the electric fields~\cite{mukamelbook}.
The three response functions in Eq.~\ref{spe} 
correspond to three different pathways -- 
(i) Stimulated Emission (SE), $R_\text{SE}$
(ii) Ground State Bleaching (GSB), $R_\text{GSB}$, 
and (iii) Excited State Absorption (ESA), $R^{*}_\text{ESA}$.
The polarization contribution from 
each pathway has sign $(-1)^{n}$,
where $n$ is the number of bra-side 
field-matter interactions.  
In this case, the contribution from 
the ESA pathway is negative whereas 
the other two are positive.


\section{The SE Pathway} \label{SE}

\begin{figure}[h!]
\begin{center}
  \includegraphics[scale=0.38]{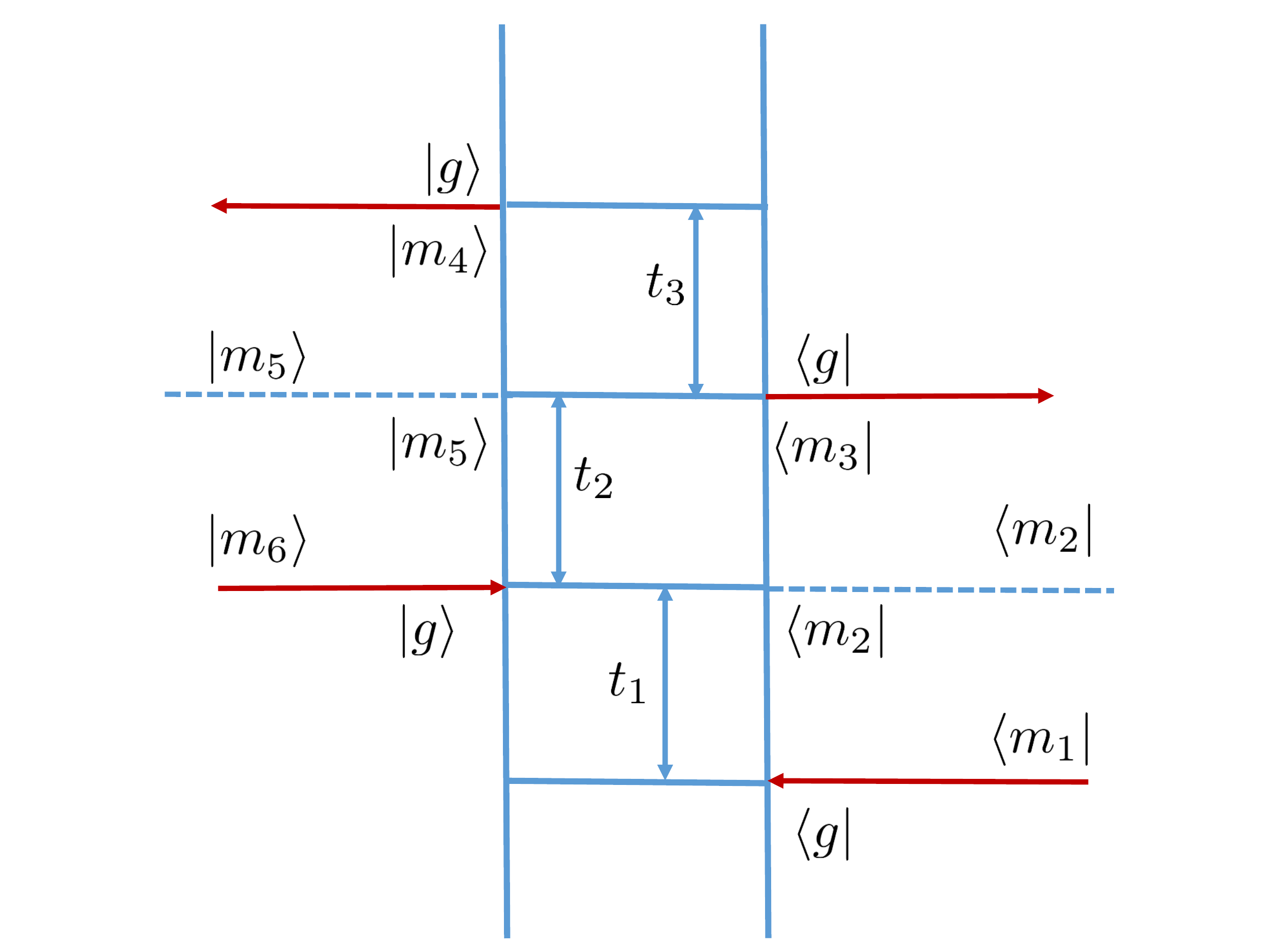}
  \caption{Stimulated Emission (SE) pathway 
  where the system-field interactions are shown 
  with red arrows.}
  \label{ese6}
  \end{center}
\end{figure}

We introduce our formulation in the context 
of the SE pathway, diagrammatically represented in 
Fig.~\ref{ese6}. The response function corresponding
to this pathway can be written as~\citep{jpcbcho}
\begin{align}
R_\text{SE}(t_{3},t_{2},t_{1}) & =\langle \mu(0) 
\mu(t_{1}+t_{2}) \mu(t_{1}+t_{2}+t_{3})\nonumber \\ 
& \times \mu(t_{1}) \rho(0) \rangle, 
\label{SEresponse0}
\end{align}
where $\rho(0)=|g\rangle\langle g|\rho_{ph}(0)$. 
Writing the time evolved transition dipole moment 
operator as 
${\mu(t)=e^{i \bar{H} t}  \mu e^{-i \bar{H} t}}$, 
we obtain,
\begin{align}
R_\text{SE}(t_{3},t_{2},t_{1})& =
Tr_{ph}\big\{\langle g |  
\mu e^{i \bar{H} t_{1}}e^{i\bar{H} t_{2}} 
\mu e^{i \bar{H} t_{3}} \mu \nonumber \\ 
& \times e^{-i \bar{H} t_{3}}e^{-i \bar{H} t_{2}} 
\mu e^{-i \bar{H} t_{1}}|g\rangle\rho_\text{ph}(0)\big\}. 
\label{SEresponse1}
\end{align}
Introducing a complete set of adiabatic
states and then unitary 
transforming to the stationary
basis, $|m \rangle$, as defined in 
Eq.~\ref{utransf}, we obtain
\begin{align}
 R_\text{SE}&(t_{3},t_{2},t_{1}) =
\sum_{\{m\}}Tr_{ph}\big\{\langle g|\mu|m_{1}\rangle
\langle m_{1} | e^{i H t_{1}} | m_{2} \rangle \nonumber \\
& \times \langle m_{2} | e^{i H t_{2}} | m_{3} \rangle 
 \langle m_{3} |  \mu | g \rangle 
 \langle g | e^{i H t_{3}} | g \rangle  
 \langle g |\mu | m_{4} \rangle \nonumber\\
& \times \langle m_{4} | e^{-i H t_{3}} | m_{5} \rangle 
\langle m_{5} | e^{-i H t_{2}} | m_{6} \rangle 
\langle m_{6} | \mu | g \rangle \nonumber \\
& \times \langle g | e^{-i H t_{1}} | g \rangle 
\rho_\text{ph} (0) \big\}, 
\label{SEresponse2}
\end{align}
where we use the index $\{ m \}$ in the summation
to denote a full sum over the set of states 
$\{m_{1},\ m_{2},\ m_{3},\ m_{4},\ m_{5},\ m_{6}\}$.
Furthermore, in Eq. \ref{SEresponse2}, since the 
transition dipole matrix element between the ground 
and an adiabatic excited electronic state can be 
reasonably assumed to be independent of $\textbf{Q}$,
we make the approximation,
$\langle g |  \mu | m_{1} (\textbf{Q}) 
\rangle \approx \langle g |  \mu | m_{1} \rangle$.
Extracting the transition dipole matrix 
elements from the phonon 
trace in Eq.~\ref{SEresponse2},
we can write
\begin{align}
&R_\text{SE}(t_{3},t_{2},t_{1})=
\sum_{\{m\}} \mu_{g m_{1}} \mu_{m_{3}g} 
\mu_{g m_{4}} \mu_{m_{6} g}\nonumber\\
& \times e^{i \epsilon_{g} (t_{3}-t_{1})}
Tr_{ph}\big\{\langle m_{1} | 
e^{i H t_{1}} | m_{2} \rangle 
\langle m_{2} | e^{i H t_{2}} | m_{3} \rangle  
\nonumber \\
&\times \langle m_{4} | e^{-i H t_{3}} | m_{5} \rangle 
\langle m_{5} | e^{-i H t_{2}} | m_{6} \rangle  
e^{i H_\text{ph}t_{3}} \nonumber \\
& \times e^{-i H_\text{ph} t_{1}} 
\rho_\text{ph} (0) \big\},
\label{SEresponse4}
\end{align}
and we have used 
$\langle g | e^{i H t} | g \rangle
=e^{i \epsilon_{g} t} e^{i H_\text{ph} t}$. 

To evaluate individual matrix elements
in Eq.~\ref{SEresponse4}, we introduce
a complete set of states in the stationary
adiabatic basis and the identity operator, $\hat{1}=e^{-iH_0t}e^{iH_0t}$,
to obtain 
\begin{align}
\langle m_{1}|e^{i Ht_{1}}|m_{2}\rangle & =
\langle m_{1} | e^{i H t_{1}} 
e^{-i H_{0} t_{1}}| m_{2} \rangle \nonumber \\ 
& \times \langle m_{2}|e^{i H_{0}t_{1}}|m_{2}\rangle. 
\label{SEresponse9}
\end{align}
Using the definition of the 
time evolution operator in the 
interaction picture,
\begin{equation}
U_{I}(t)=e^{iH_{0}t} e^{-iHt},
\label{eq:int_time}
\end{equation}
and evaluating time evolution
under the zeroth order Hamiltonian we obtain,
\begin{equation}
e^{-i H_{0} t}|m \rangle=
\left(\hat{T}e^{-i\int^{t}_{0}dt'
\varepsilon(\textbf{Q}(t'))}\right)
e^{-i H_\text{ph}t}|m \rangle,
\end{equation}
where $\textbf{Q}(t')=e^{i H_\text{ph} t'}\textbf{Q} 
e^{-i H_\text{ph} t'}$ 
and $\hat{T}$ is the 
time-ordering operator~\cite{pallavi1}.
This allows us to write Eq.~\ref{SEresponse9} as  
\begin{align}
\langle m_{1}|e^{iHt_{1}}|m_{2}\rangle
& =\langle m_{1}|U_{I}^{\dag}(t_{1})|m_{2}\rangle \nonumber \\ 
& \times \left(\hat{T}^{\dag}e^{i\int^{t_{1}}_{0}dt'
\varepsilon_{m_{2}}(\textbf{Q}(t'))}\right)
e^{i H_{ph}t_{1}}. 
\label{SEresponse10}
\end{align}
Further, recognizing that the matrix elements 
of the time-evolution operator in the 
interaction picture are~\cite{tannor}
\begin{equation}
\langle m | U_{I}(t)| n \rangle=(\hat{T}e^{-i\int^{t}_{0}dt'H_{na}(t')})_{m,n}, \label{exteqn}
\end{equation}
allows us to re-write the expression 
in Eq.~\ref{SEresponse10} as
\begin{align}
\langle m_{1}|e^{i H t_{1}}|m_{2}\rangle
& =\left(\hat{T}^{\dag}
e^{i\int^{t1}_{0}dt'H_{na}(t')}\right)_{m_{1},m_{2}} \nonumber \\ 
& \times \left(\hat{T}^{\dag}e^{i\int^{t_{1}}_{0}dt'
\varepsilon_{m_{2}}(\textbf{Q}(t'))}\right)
e^{i H_{ph}t_{1}},
\label{SEresponse11}
\end{align}
where $H_{na}$ is previously defined 
in Eq.~\ref{Hna},  and 
$H_{na}(t)~=~e^{i H_{0}(t)}H_{na}e^{-i H_{0}(t)}$.

The terms in Eq.~\ref{SEresponse11}
yield significant physical insight.
Terms in the exponent that contain $H_{na}$ 
depend on momenta and give rise to nonadiabatic 
transitions that cause population relaxation.
The remaining terms in the exponent depend
on bath position coordinates, 
$\varepsilon_{m}(\textbf{Q}(t'))$), 
and account for environment driven 
fluctuations in the energies of the 
stationary states and cause decoherence. 

Taylor expanding $\varepsilon_m(\textbf{Q})$ about 
$\textbf{Q}=0$ to first order, we obtain
\begin{align}
\hat{T}e^{-i\int^{t}_{0}dt'
\varepsilon_{m}(\textbf{Q}(t'))} & =
e^{-i \varepsilon_{m}t} \nonumber \\
& \times \left(\hat{T}e^{-i\int^{t}_{0}dt'\nabla_{\textbf{Q}} 
\varepsilon_{m}(\textbf{Q}(t'))
\cdot\textbf{Q}(t')}\right),
\label{eq:timeorder}
\end{align}
where 
$\nabla_{\textbf{Q}} \varepsilon_{m}(\textbf{Q}(t'))$ 
is the gradient.
Substituting Eq.~\ref{SEresponse11} 
and Eq.~\ref{eq:timeorder} into the 
product of matrix elements in 
Eq.~\ref{SEresponse4}, we obtain
\begin{align}
R_\text{SE}&(t_{3},t_{2},t_{1}) =
\sum_{\{m\}} \mu_{g m_{1}} \mu_{m_{3}g} 
\mu_{g m_{4}} \mu_{m_{6} g} \nonumber \\
& \times e^{i (\varepsilon_{m_{2}}-\epsilon_{g}) t_{1}} 
e^{i (\varepsilon_{m_{3}}-\varepsilon_{m_{5}}) t_{2}}  
e^{-i (\varepsilon_{m_{4}}-\epsilon_{g}) t_{3}} \nonumber \\
& 
\times F_\text{SE}(t_{1},t_{2},t_{3}),
\label{SEresponse12}
\end{align}
where the pre-exponential factor is defined as
\begin{align}
& F_\text{SE}(t_1,t_2,t_3) =
\big\langle \left(1-D_\text{SE}(t_{1},t_{2},t_{3})\right) \nonumber \\
& \times \left(\delta_{m_{1},m_{2}}
\delta_{m_{2},m_{3}}\delta_{m_{4},m_{5}}
\delta_{m_{5},m_{6}} -P_\text{SE}(t_{1},t_{2},t_{3})\right)\big\rangle, \label{eq:preexp}
\end{align}
and we use $\langle\cdot\rangle$ as short-hand for
the phonon trace $\text{Tr}\{\ldots\rho_{ph}(0)\}$.
In Eq.~\ref{eq:preexp}, 
the decoherence term is defined as 
\begin{align}
D_\text{SE}(t_{1},&t_{2},t_{3})=
1-\left((\hat{T}^{\dag}e^{i\int^{t_{1}}_{0}dt'\nabla_{\textbf{Q}} 
\varepsilon_{m_{2}}(\textbf{Q}(t'))\cdot \textbf{Q}(t')})\right.\nonumber \\
& \times (\hat{T}^{\dag}e^{i\int^{t_{2}}_{0}dt'\nabla_{\textbf{Q}} 
\varepsilon_{m_{3}}(\textbf{Q}(t'))\cdot \textbf{Q}(t')})\nonumber \\
& \times (\hat{T}e^{-i\int^{t_{3}}_{0}dt'\nabla_{\textbf{Q}} 
\varepsilon_{m_{4}}(\textbf{Q}(t'))\cdot \textbf{Q}(t')})\nonumber \\ 
& \left.\times (\hat{T}e^{-i\int^{t_{2}}_{0}dt'\nabla_{\textbf{Q}} 
\varepsilon_{m_{5}}(\textbf{Q}(t'))\cdot \textbf{Q}(t')})\right), 
\label{SEresponse14}
\end{align}
and the population relaxation term is defined as
\begin{align}
P_\text{SE}(t_{1},t_{2},t_{3})
&= \delta_{m_{1},m_{2}}\delta_{m_{2},m_{3}}\delta_{m_{4},m_{5}}
\delta_{m_{5},m_{6}} \nonumber \\ 
&-\left(\left(\hat{T}^{\dag}e^{i\int^{t_{1}}_{0}dt'H_{na}(t')}
\right)_{m_{1},m_{2}}\right. \nonumber \\
&\times \left(\hat{T}^{\dag}e^{i\int^{t_{2}}_{0}dt'H_{na}(t')}
\right)_{m_{2},m_{3}} \nonumber \\ 
&\times \left(\hat{T}e^{-i\int^{t_{3}}_{0}dt'H_{na}(t')}
\right)_{m_{4},m_{5}} \nonumber \\
&\left.\times \left(\hat{T}e^{-i\int^{t_{2}}_{0}dt'H_{na}(t')}
\right)_{m_{5},m_{6}} \right).
\label{SEresponse15}
\end{align}
It is worth noting that the expressions for $D_\text{SE}(t_{1},t_{2},t_{3})$ and $P_\text{SE}(t_{1},t_{2},t_{3})$ are dependent on the values of $m_{1}$, $m_{2}$, $m_{3}$, $m_{4}$, $m_{5}$ and  $m_{6}$, respectively but the dependence is not explicitly stated in $D_\text{SE}(t_{1},t_{2},t_{3})$ and $P_\text{SE}(t_{1},t_{2},t_{3})$ to avoid cluttering.


\subsection{Cumulant Expansion} \label{cumulant}

The decoherence and population relaxation terms in 
Eq.~\ref{SEresponse14} and Eq.~\ref{SEresponse15} respectively
are evaluated using a second order cumulant expansion~\cite{mukamelbook}. 
While one of us has previously used this approach 
to evaluate the decoherence term~\cite{pallavi1, pallavi2}, 
here we propose a cumulant expansion approach to treat
the population relaxation term as well, eliminating 
the need for master equation based methods.
A significant benefit of this approach is its 
computational efficiency: the decoherence term can 
be evaluated analytically for Ohmic and Debye spectral 
density functions and the population relaxation term 
can be calculated using simple numerical integration.

We consider two cases in evaluating Eq.~\ref{eq:preexp}:\\
Case 1: 
$\delta_{m_{1},m_{2}}\delta_{m_{2},m_{3}}\delta_{m_{4},m_{5}}
\delta_{m_{5},m_{6}}=1$, where the conditions $m_{1}=m_{2}=m_{3}$ 
and $m_{4}=m_{5}=m_{6}$ are both satisfied. 
Neglecting the coupling between the decoherence and population 
relaxation term and using a second order cumulant expansion
we obtain,
\begin{align}
F_\text{SE}(&t_1,t_2,t_3) \nonumber \\
& \approx 
1-\big\langle D_\text{SE}(t_1,t_2,t_3)\big\rangle 
-\big\langle P_\text{SE}(t_1,t_2,t_3)\big\rangle \nonumber \\
& \approx e^{-\big( \big\langle \bar{D}_\text{SE}(t_{1},t_{2},t_{3})
\big\rangle+\big\langle \bar{P}_\text{SE}(t_{1},t_{2},t_{3})\big\rangle \big)},
\label{SEresponse19}
\end{align}
where $\bar{D}_\text{SE}(t_{1},t_{2},t_{3})$ and 
$\bar{P}_\text{SE}(t_{1},t_{2},t_{3}))$ involve
a series of single and double time integrals
detailed in the appendix A and B respectively. 
We note that $\bar{D}_\text{SE}$ is analytically
determined for thermal baths described by Ohmic or 
Debye spectral densities 
and can be numerically evaluated 
for a general spectral density
function.\\ 
Case 2: 
$\delta_{m_{1},m_{2}}\delta_{m_{2},m_{3}}\delta_{m_{4},m_{5}}
\delta_{m_{5},m_{6}}=0$. As in the previous case, neglecting 
the coupling between decoherence and population relaxation
\begin{multline}
F_\text{SE}(t_1,t_2,t_3)\approx \big\langle -P_\text{SE}\big\rangle  
=\big\langle 1-\left(1+P_\text{SE}\right)\big\rangle . 
\label{SEresponse21}
\end{multline}
Using a second order cumulant expansion, we 
obtain
\begin{multline}
F_\text{SE}(t_{1},t_{2},t_{3}) \approx 
1-e^{\big\langle\bar{P}_\text{SE}(t_{1},t_{2},t_{3})\big\rangle},
\label{SEresponse23}
\end{multline}
where $\bar{P}_\text{SE}$ is defined in the Appendix B.
\section{GSB and ESA Pathways}\label{othertwo}
\begin{figure}[tpbh]
\vspace{-0.2in}
\begin{center}
  \includegraphics[scale=0.38]{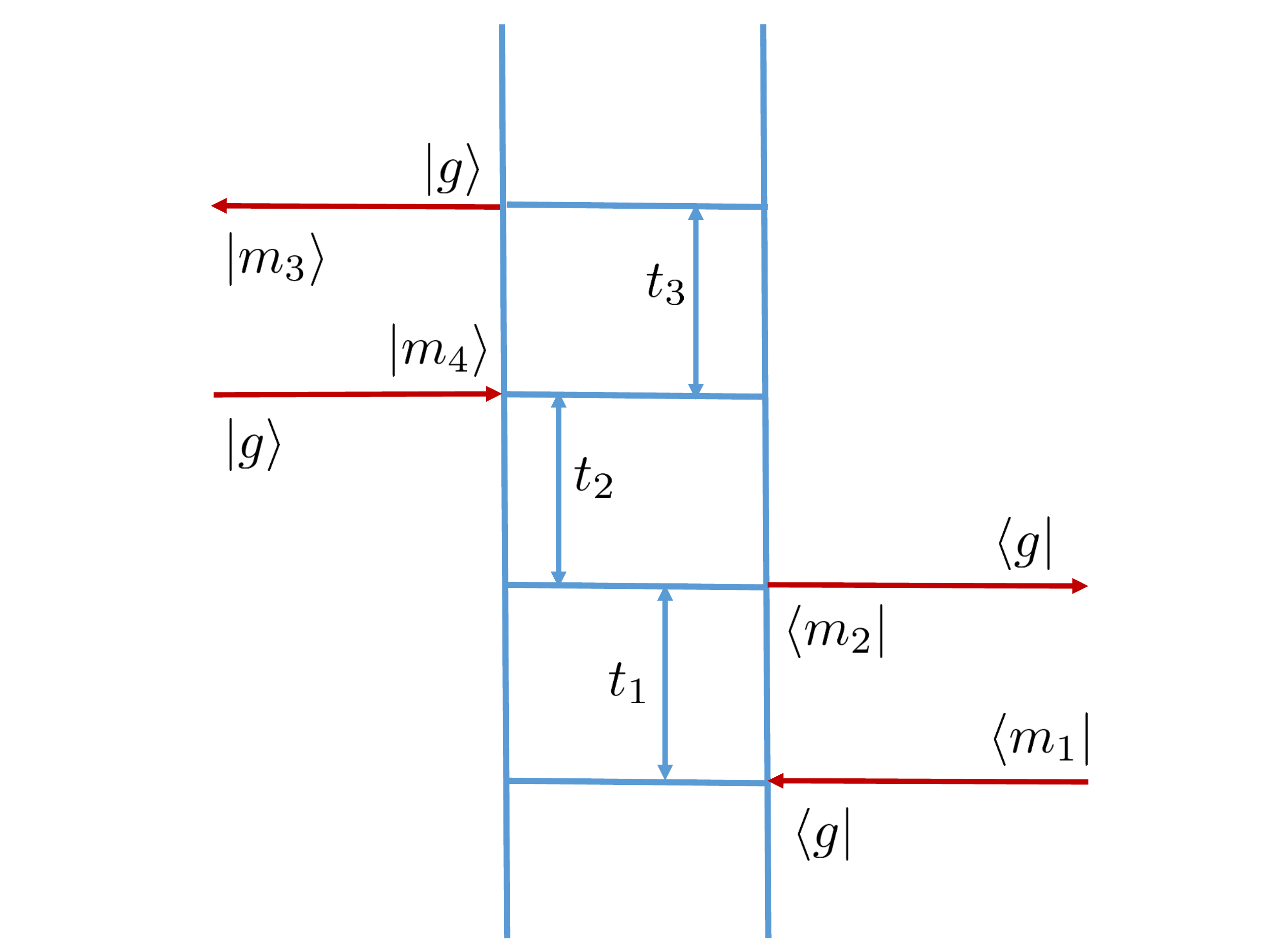}
  \caption{The GSB pathway where 
  system field interactions are indicated by
  red arrows}
  \label{gsbrf}
  \end{center}
\vspace{-0.2in}
\end{figure}
The response function for the Ground State Bleaching (GSB) 
pathway shown in Fig.~\ref{gsbrf} is~\citep{jpcbcho}
\begin{align}
R_\text{GSB}(t_{3},t_{2},t_{1})&=
\langle \mu(0) \mu(t_{1}) \mu(t_{1}+t_{2}+t_{3})\nonumber  \\ 
& \times \mu(t_{1}+t_{2})\rho_{0} \rangle. 
\label{GSBRF1}
\end{align}   
Extracting the transition dipole matrix elements 
and introducing complete sets of stationary states,
we obtain
\begin{align}
& R_\text{GSB}(t_{3},t_{2},t_{1})=\sum_{\{m\}}\mu_{gm_{1}} 
\mu_{m_{2}g} \mu_{gm_{3}} \mu_{m_{4}g}
\nonumber\\ 
&\times Tr_{ph}\big\{\rho_{ph}(0)
\langle m_{1} |e^{i \bar{H} t_{1}} |m_{2} \rangle 
\langle g| e^{i \bar{H} (t_{2}+t_{3})} |g \rangle \nonumber\\
& \times \langle m_{3} | e^{-i \bar{H} t_{3}} | m_{4} \rangle
\langle g | e^{-i \bar{H}(t_{1}+t_{2})} |g \rangle \big\}. 
\label{GSBRF3}
\end{align}
Evaluating the matrix elements in Eq.~\ref{GSBRF3}, 
and using a second order cumulant expansion to 
approximate the decoherence and population relaxation 
terms, we arrive at an easily evaluated expression 
for the response function. 
The final expression along with the derivation details are 
provided in Appendix C.\\
\begin{figure}[tpbh]
\vspace{-0.15in}
\begin{center}
  \includegraphics[scale=0.38]{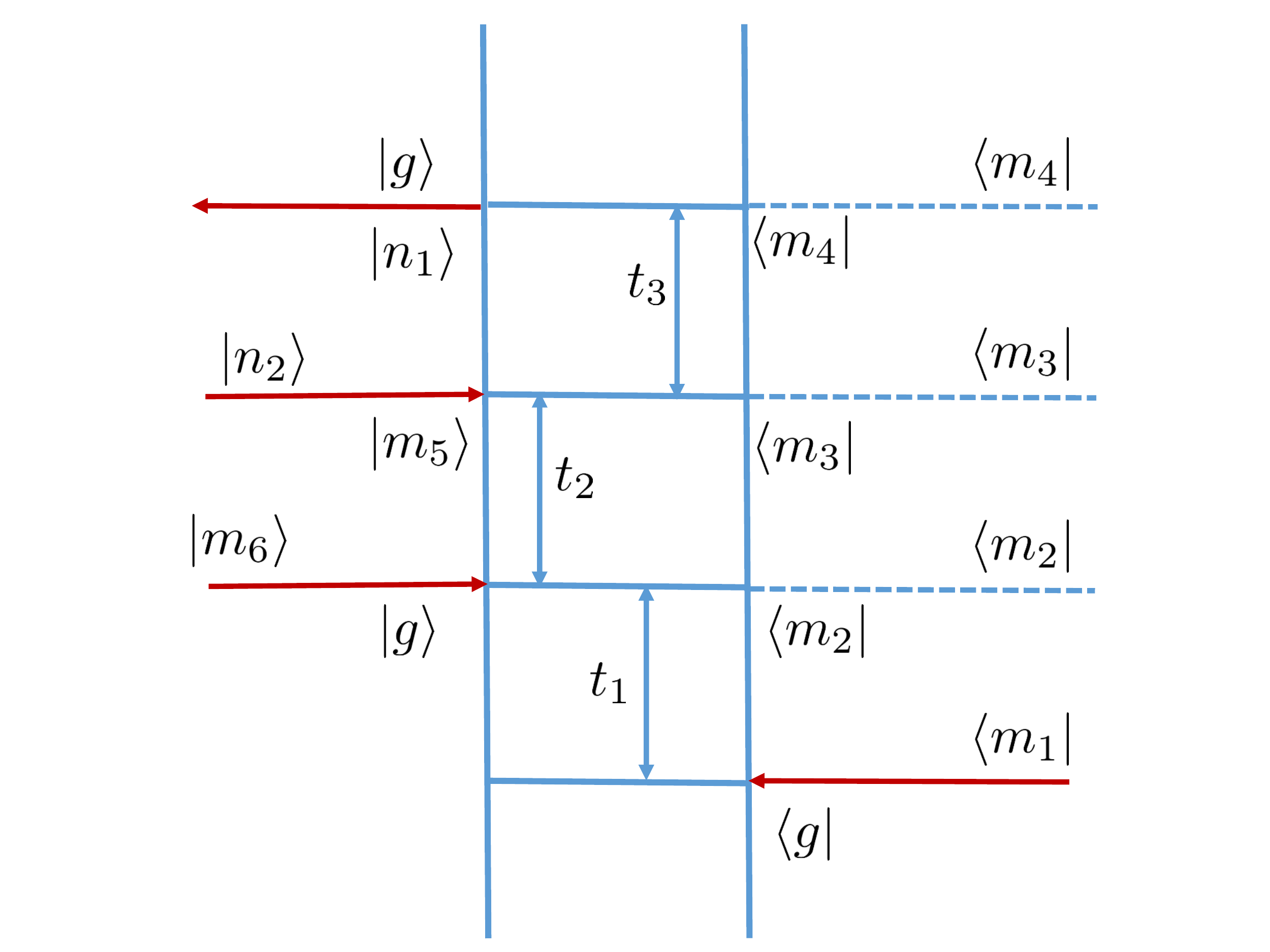}
  \caption{The ESA pathway where 
  system field interactions are indicated by
  red arrows}
  \label{esarf}
  \end{center}
\vspace{-0.2in}
\end{figure}
The Excited State Absorption (ESA) pathway involves 
both the singly and doubly excited states. 
Local doubly excited state are represented as 
$|i,j \rangle$, with the condition $i<j$ to 
avoid double counting of states. 
The electronic coupling between a pair of doubly 
excited states is given as 
$\langle i,j|\bar{H}|i_{1},j_{2} \rangle =J_{i,j_{2}}$ 
if $j=i_{1}$ and 
$\langle i,j|\bar{H}|i_{1},j_{2} \rangle =J_{j,j_{2}}$ 
if $i=i_{1}$ and zero otherwise.
The response function for the ESA pathway shown 
in Fig.~\ref{esarf} is then given as~\citep{jpcbcho}
\begin{align}
R^{*}_\text{ESA}(t_{3},t_{2},t_{1})& = 
\langle \mu(0) \mu(t_{1}+t_{2}+t_{3})\nonumber \\ 
& \times \mu(t_{1}+t_{2}) \mu(t_{1})\rho_{0} \rangle. 
\label{ESARF1}
\end{align}
As  before, we extract the transition dipole 
matrix elements from the trace and introduce
the $|m \rangle$ stationary states obtained 
by unitary transforming singly excited adiabatic states 
and $|n \rangle$ stationary states obtained 
by unitary transforming doubly excited adiabatic states.
\begin{align}
&R^{*}_\text{ESA}(t_{3},t_{2},t_{1})= 
\sum_{\{m,n\}}\mu_{gm_{1}} \mu_{m_{4}n_{1}} 
\mu_{n_{2}m_{5}} \mu_{m_{6}g} \nonumber\\ 
& \times Tr_{ph}\big\{
\rho_{ph}(0) \langle m_{1}|e^{i \bar{H} t_{1}} |m_{2} \rangle 
\langle m_{2}|e^{i \bar{H} t_{2}} |m_{3} \rangle 
\nonumber \\
& \times \langle m_{3} | e^{i \bar{H} t_{3}}|m_{4} \rangle 
\langle n_{1}|e^{-i \bar{H} t_{3}}|n_{2}\rangle 
 \langle m_{5} | e^{-i \bar{H} t_{2}}|m_{6} \rangle \nonumber \\
& \times \langle g | e^{-i \bar{H} t_{1}} | g \rangle\big\}. 
\label{ESARF3}
\end{align}

We note that transitions from the singly excited states, 
$|m_{i} \rangle$, to the doubly excited states, $|n_{i} \rangle$,
are only induced by the applied electric field and not via
phonon-mediated population relaxation. 
The final expression for the ESA response function 
and derivation details for the same are provided in Appendix D.\\*

\section{Results and Discussion} \label{rdsection}

\begin{figure*}
\begin{center}
  \includegraphics[scale=0.52]{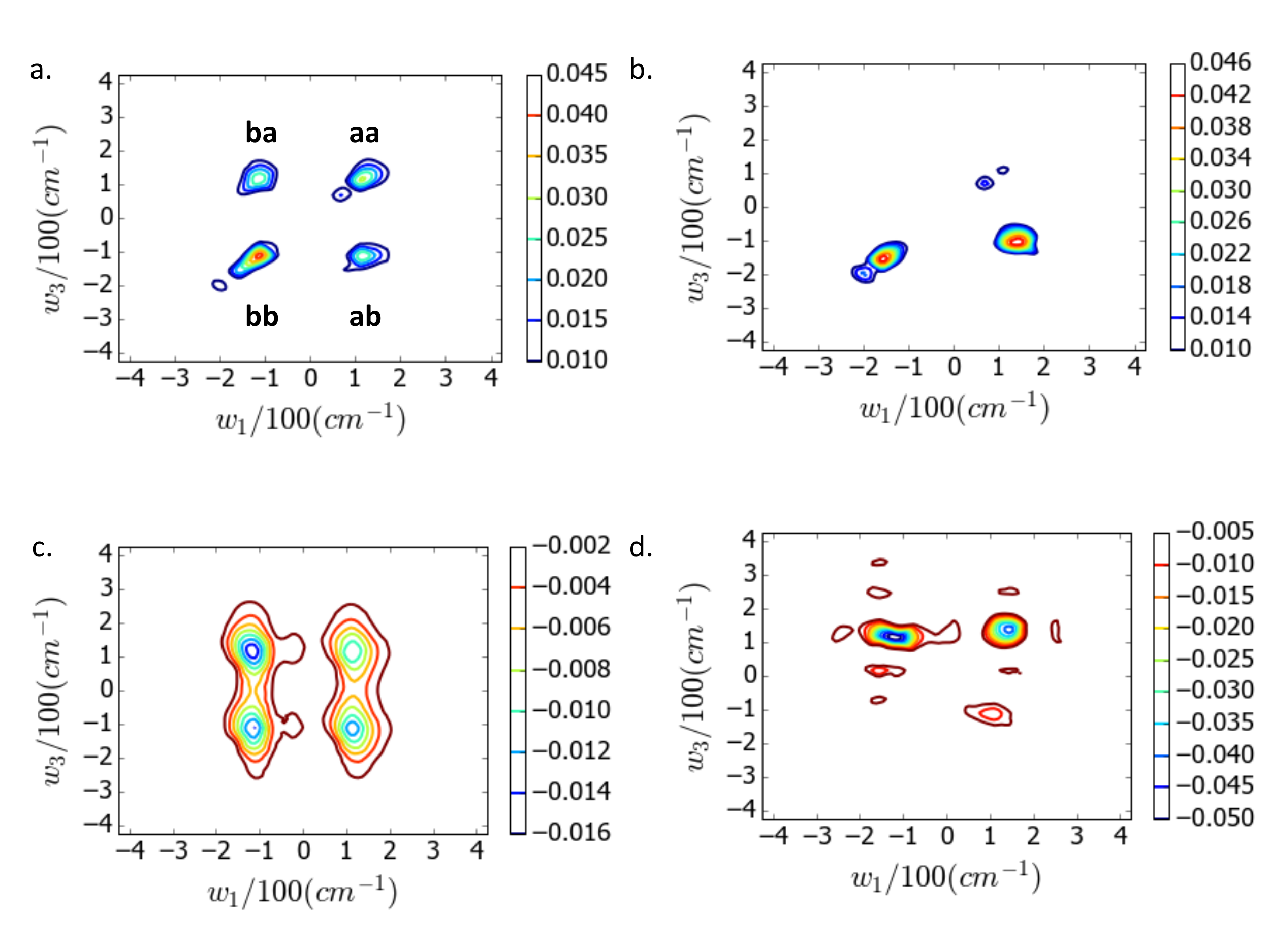} 
  \caption{Contribution to the response function from the SE and ESA pathways at different $t_{2}$: (a) SE contribution at $t_{2}=10\ fs$, (b) SE contribution at $t_{2}=625\ fs$, (c) ESA contribution at $t_{2}=10\ fs$, and (d) ESA contribution at $t_{2}=625\ fs$.} \label{all}
  \end{center}
\end{figure*}

\begin{figure}
\begin{center}
  \includegraphics[scale=0.3]{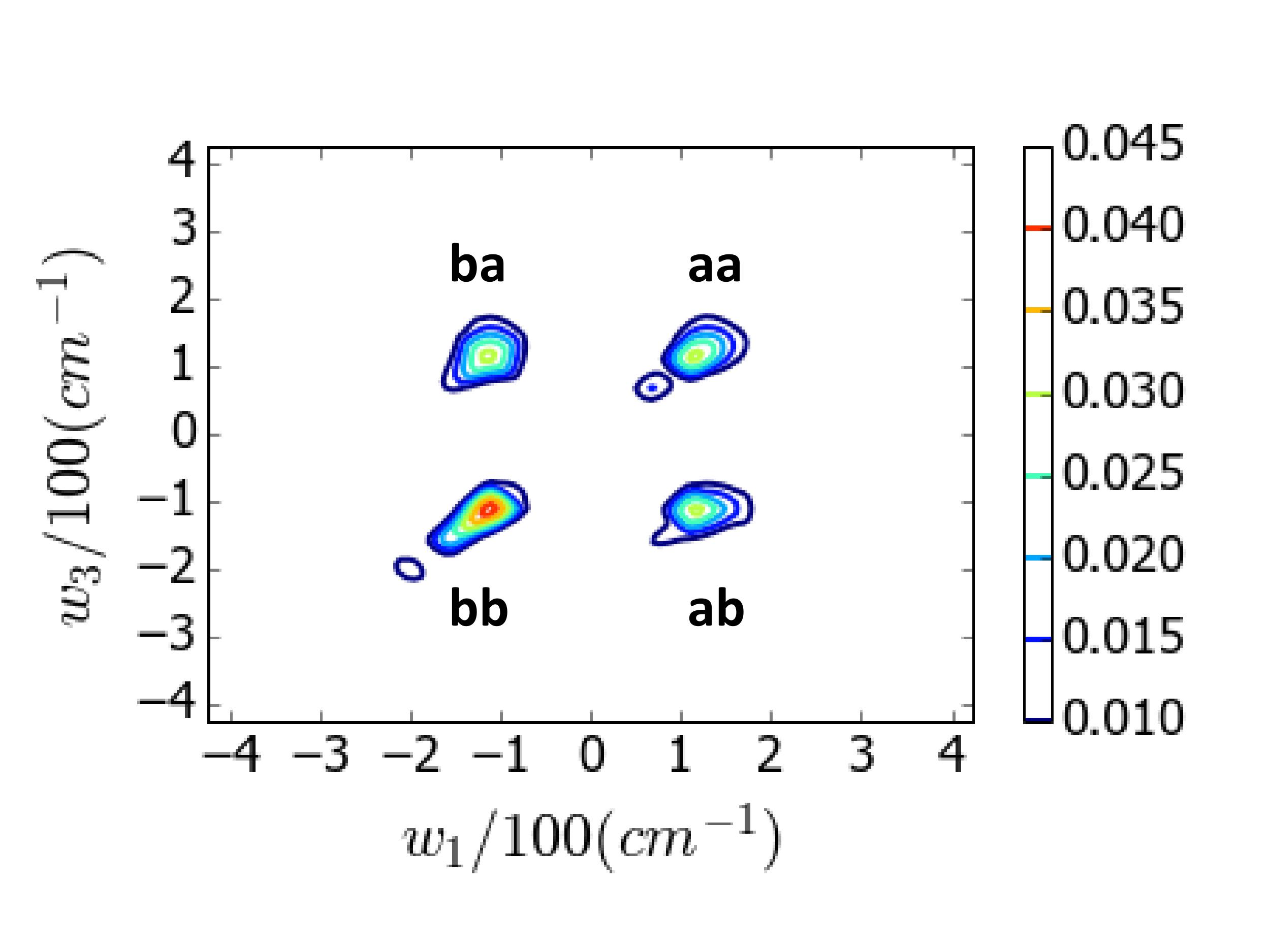} 
  \caption{Contribution to the response function from the GSB pathway at $t_{2}=10\ fs$.} \label{gsb10}
  \end{center}
\end{figure}

We calculate the 2D electronic spectrum for a two-level
system where the Hamiltonian $\bar{H}$ (see Eq.~\ref{Hamiltonian}) 
is given as,
\begin{align}
\bar{H} & =\epsilon_{g}| g \rangle \langle g |+ \epsilon_{1} | 1 \rangle \langle 1 | +  \epsilon_{2} | 2 \rangle \langle 2 |+
J_{12} |1\rangle \langle 2 | \nonumber \\ 
&+J_{21} |2\rangle \langle 1 | +Q_{1} |1 \rangle \langle 1|+Q_{2} |2 \rangle \langle 2|+H_{ph} ,\nonumber \\
\end{align}
where $\epsilon_{g}=-12000\ cm^{-1}$, $\epsilon_{1}=-50\ cm^{-1}$, $\epsilon_{2}=50\ cm^{-1}$ and $J_{12}=J_{21}=100\ cm^{-1}$. Transforming to the stationary basis (see Eq.~\ref{H0}), we have states $a$ and $b$ with energies $\varepsilon_{a}(\textbf{Q}=\mathbf{0})=111.803\ cm^{-1}$ and $\varepsilon_{b}(\textbf{Q}=\mathbf{0})=-111.803\ cm^{-1}$. The thermal bath (environment) is modeled by an 
Ohmic spectral density, given as 
\begin{equation}
S(\omega)=\frac{\lambda}{\omega_{c}}\omega 
e^{-\omega/\omega_{c}}, 
\label{rd}
\end{equation}
where $\lambda$ is the reorganization energy 
and $\omega_{c}$ is the phonon relaxation frequency. 
We use the values $\frac{\lambda}{\omega_{c}}=1.2$, 
$\omega_{c}=53\ cm^{-1}$, and temperature $T=77\ K$. 
We further assume that the transition dipole 
$\mu_{ga}=\mu_{gb}$. 

The three pathways contribute different spectral features 
to the overall 2D spectrum at different times.
The diagonal peaks in the spectrum are labeled, 
$\bf{aa}$ centered at $\omega_1=\omega_3=\omega_a$ 
and $\bf{bb}$ centered at $\omega_1=\omega_3=\omega_b$, 
and the off-diagonal peaks are labeled, 
$\bf{ab}$ centered at $\omega_1=\omega_a\;,\omega_3=\omega_b$ 
and $\bf{ba}$ centered at $\omega_1=\omega_b\;,\omega_3=\omega_a$.

For the SE pathway, at $t_{2}=0$, the populations are centered at the diagonal peaks $\bf{aa}$ and $\bf{bb}$, respectively and the coherences are centered at the off-diagonal peaks $\bf{ab}$ and $\bf{ba}$, respectively. Fig. \ref{all}(a) shows the contributions from the populations (peaks $\bf{aa}$ and $\bf{bb}$) and coherences (peaks $\bf{ab}$ and $\bf{ba}$) at a short time $t_{2}=10\ fs$. It is to be noted that at short times, phonon-mediated population transfer is insignificant as phonons are not thermally activated yet. However, at longer times, the thermally activated phonons result in a population transfer from $a$ to $b$, resulting in an emerging off-diagonal peak at $\bf{ab}$ and decreasing intensity at $\bf{aa}$. Similarly, we will have population relaxation from $b$ to $a$, leading to an off-diagonal peak at $\bf{ba}$ and decreasing intensity at $\bf{bb}$. Again, the rate of downhill relaxation ($a \rightarrow b$) is greater than that of the uphill relaxation pathway ($b \rightarrow a$), resulting in a larger intensity at $\bf{ab}$ compared to $\bf{ba}$. Decoherence, on the other hand, leads to decreasing contributions from coherences at the peaks $\bf{ab}$ and $\bf{ba}$ with increasing $t_{2}$. Fig. \ref{all}(b) shows the SE pathway contributions from both populations and coherences at $t_{2}=625\ fs$. Decoherence is effectively complete and population relaxation, as discussed above, leads to large intensities at the peaks $\bf{bb}$ and $\bf{ab}$, respectively and a decrease in intensity at $\bf{aa}$. At $77\ K$, thermal energy is insufficient to access the uphill pathway $b \rightarrow a$, hence there is effectively no peak due to population relaxation at $\bf{ba}$.

The ESA pathway, at $t_{2}=0$, results in off-diagonal 
peaks, $\bf{ab}$ and $\bf{ba}$ for populations at $a$ and at $b$, respectively
and coherences at peaks $\bf{bb}$ and $\bf{aa}$, respectively.
As discussed before, the intensity contributions from the 
ESA pathway are negative. Fig. \ref{all}(c) shows the populations at off-diagonal peaks and coherences at diagonal peaks at $t_{2}=10\ fs$. At longer times, decoherence will result in decreasing contributions from coherences and downhill population relaxation ($a \rightarrow b$) will result in a decreased contribution at $\bf{ab}$ and a rise in negative intensity at $\bf{aa}$. The negative intensity at $\bf{ba}$, on the other hand, does not change much since uphill population relaxation is insignificant at $T=77\ K$. These features are seen in Fig. \ref{all}(d), which shows the ESA pathway contributions at $t_{2}=625\ fs$.

In the GSB pathway, there is only ground state 
dynamics during $t_{2}$. 
Hence, the excited state populations at the diagonal 
peaks and coherences at the off-diagonal peaks 
do not evolve with $t_{2}$ (see Fig. \ref{gsb10}).

\begin{figure*}
\begin{center}
  \includegraphics[scale=0.52]{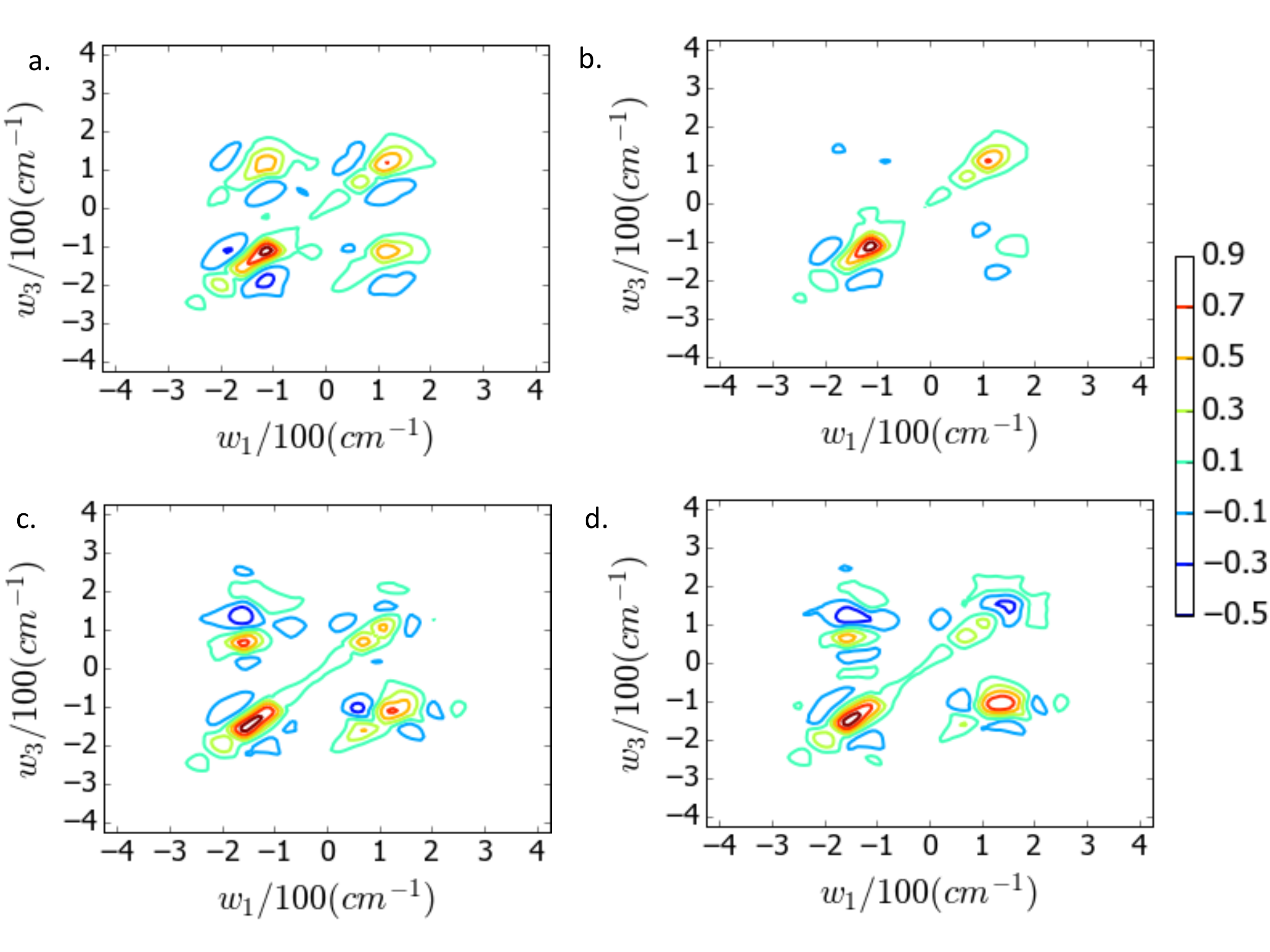} 
  \caption{2D photon echo electronic spectra at different $t_{2}$: (a) $10\ fs$, (b) $100\ fs$, (c) $300\ fs$ and (d) $625\ fs$.} \label{all1}
  \end{center}
\end{figure*}

The overall spectrum arising from the contributions of the 
SE, GSB and ESA pathways are shown for different times in 
Figs. \ref{all1}(a)-(d). Immediate and marked differences could be spotted at short 
and long $t_{2}$, respectively. 
At short $t_{2}$ (see Fig. \ref{all1}(a), $t_{2}=10\ fs$), we have 
positive intensities arising mostly from populations at 
the peaks $\bf{aa}$ and $\bf{bb}$ and coherences at the off-diagonal peaks $\bf{ab}$ and $\bf{ba}$. 
Fig. \ref{all1}(b) shows the spectrum at $t_{2}=100\ fs$. Decoherence is complete and a small peak is seen emerging at $\bf{ab}$, due to the downhill population relaxation $a \rightarrow b$ in SE pathway. Fig. \ref{all1}(c) shows the spectrum at $t_{2}=300\ fs$. The crosspeak at $\bf{ab}$ has increased in intensity, with a concomitant decrease in intensity at $\bf{aa}$. A negative intensity is also seen at $\bf{ba}$, arising from the ESA pathway due to population at $b$. Fig. \ref{all1}(d) shows the spectrum at $t_{2}=625\ fs$. The intensities at $\bf{bb}$ and $\bf{ab}$ (due to population relaxation from $a$ to $b$) 
are large and positive, whereas negative intensities, 
arising from the ESA pathway are seen at peaks $\bf{ba}$ (large negative intensity due to population at $b$) and $\bf{aa}$ (due to $a \rightarrow b$ relaxation). Also, decoherence is complete.

\section{Conclusions}

We introduce a new method for simulating electronic 2DPES (2D Photon Echo Spectroscopy). We transform to  a stationary basis and employ the cumulant expansion approach to evaluate decoherence and population relaxation. We demonstrate the efficiency of our new approach for a model two level system. We capture all the features expected from 2DPES, the most prominent of them being an emerging off-diagonal peak (at $\bf{ab}$) at long $t_{2}$ from the SE pathway due to population relaxation from the higher to the lower energy exciton, as well as a negative off-diagonal peak (at $\bf{ba}$) arising from the ESA pathway. The coherences decay with time and have oscillatory behavior, in contrast to exciton population relaxation, which is reflected by a steady decrease/increase in the relevant peaks. We will leverage the computational efficiency of our approach to simulate energy transfer in higher dimensional systems.

\section{Acknowledgements}
The authors acknowledge the Cornell startup funding and 
DOE NMGC seed funding. 
The authors are also grateful to Professor K. L. Sebastian 
for helpful discussions.

\bibliography{draft10}

\vspace{10 mm}

\pagebreak
\begin{center}
\textbf{\large Appendices}
\end{center}
\setcounter{equation}{0}
\setcounter{figure}{0}
\setcounter{table}{0}
\setcounter{page}{1}
\makeatletter
\renewcommand{\theequation}{S\arabic{equation}}
\renewcommand{\thefigure}{S\arabic{figure}}
\appendix
\section*{Appendix A: Decoherence in the Stimulated Emission Pathway} 
\label{app_deco}

$\bar{D}_\text{SE}(t_{1},t_{2},t_{3})$, when written out in full, 
contains several second order time-ordered terms. These involve 
integration with respect to two different time arguments, $t$ 
and $s$, over an integrand, which, when traced over, has 
the general form,
\begin{align}
d_{m,n}(t,s)&=\big\langle (\nabla_{\textbf{Q}} 
\varepsilon_{m}(\textbf{Q}(t))\cdot \textbf{Q}(t)) \nonumber  \\
&\times (\nabla_{\textbf{Q}} \varepsilon_{n}(\textbf{Q}(s))
\cdot \textbf{Q}(s)) \big\rangle.
\label{decoherence2}
\end{align}
We neglect the minimal contribution from the higher 
order derivatives 
$\frac{\partial^{n} \varepsilon_{m}}{\partial Q_{j}^{n}}$, 
where $n>1$ and $j$ labels the site/chromophore~\cite{pallavi2}. 
For an uncorrelated bath, this gives
\begin{equation}
d_{m,n}(t,s)=\sum_{j}\big(\frac{\partial \varepsilon_{m}}
{\partial Q_{j}}\big)\big(\frac{\partial \varepsilon_{n}}
{\partial Q_{j}}\big) \big\langle Q_{j}(t) Q_{j}(s) \big\rangle. 
\label{decoherence3}
\end{equation}
Noting that $Q_{j}=\sum_{b} m_{jb} \nu_{jb} q_{jb}$, 
Eq.~\ref{decoherence3} can be written as
\begin{align}
d_{m,n}(t,s)&=\int^{\infty}_{0}d\omega S(\omega)
\big(\coth(\frac{\beta \omega}{2})\nonumber \\ & \times \cos(\omega(t-s))-
i \sin (\omega (t-s))\big). 
\label{dmn}
\end{align}
Here, $S(\omega)$ is the spectral density modeling 
the environment and is defined as 
$S(\omega)=\sum_{b}\frac{m_{jb}\nu_{jb}^2}{2 \omega_{jb}}
\delta(\omega-\omega_{jb})$. 
$\bar{D}_\text{SE}(t_{1},t_{2},t_{3})$ is obtained 
analytically for the Ohmic and Debye spectral densities. 
For other specific spectral densities, this needs to 
be obtained numerically and involves at most $10$ 
numerical integrations which are easily evaluated 
(see Eq.~\ref{decoherence1}).

The SE pathway decoherence, 
$\bar{D}_\text{SE}(t_{1},t_{2},t_{3})$, 
contains 4 types of terms:

(a) a first order term, 
\begin{equation}
\int^{t}_{0}dt'(\nabla_{\textbf{Q}} \varepsilon_{m}(\textbf{Q}(t'))\cdot \textbf{Q}(t')),
\end{equation}
which when traced over, gives zero as ${\big\langle Q_{j}(t) \big\rangle=0}$,

(b) a time-ordered second order term
\begin{multline}
X_{m}(t)=\int^{t}_{0}dt'\int^{t'}_{0}dt''(\nabla_{\textbf{Q}}  \varepsilon_{m}(\textbf{Q}(t'))\cdot \textbf{Q}(t'))\\(\nabla_{\textbf{Q}}  \varepsilon_{m}(\textbf{Q}(t''))\cdot \textbf{Q}(t'')), \label{X}
\end{multline}
(c) a second order term given by a product of two first order terms with different time arguments $t$ and $s$
\begin{multline}
Y_{m,n}(t,s)=\int^{t}_{0}dt'\int^{s}_{0}dt''(\nabla_{\textbf{Q}}  \varepsilon_{m}(\textbf{Q}(t'))\cdot \textbf{Q}(t'))\\(\nabla_{\textbf{Q}}  \varepsilon_{n}(\textbf{Q}(t''))\cdot \textbf{Q}(t'')), \label{Y}
\end{multline}
(d) a second order term given by a product of two first order terms with the same time argument $t$ \begin{multline}
Z_{m,n}(t)=\int^{t}_{0}dt'\int^{t}_{0}dt''(\nabla_{\textbf{Q}}  \varepsilon_{m}(\textbf{Q}(t'))\cdot \textbf{Q}(t'))\\(\nabla_{\textbf{Q}}  \varepsilon_{n}(\textbf{Q}(t''))\cdot \textbf{Q}(t'')). \label{Z}
\end{multline}

The second order terms are evaluated analytically for the Ohmic spectral density (Eq. \ref{rd}) \cite{pallavi1, pallavi2}.

$\bar{D}_\text{SE}(t_{1},t_{2},t_{3})$ is given as,

\begin{align}
\bar{D}_\text{SE}(t_{1},t_{2},t_{3})& =X^{\dag}_{m_{2}}(t_{1})+X^{\dag}_{m_{3}}(t_{2})+X_{m_{4}}(t_{3})\nonumber \\& +X_{m_{5}}(t_{2})+Y_{m_{2},m_{3}}(t_{1},t_{2}) \nonumber \\ & -Y_{m_{2},m_{4}}(t_{1},t_{3}) -Y_{m_{2},m_{5}}(t_{1},t_{2})\nonumber \\ & -Y_{m_{3},m_{4}}(t_{2},t_{3})-Z_{m_{3},m_{5}}(t_{2}) \nonumber   \\ & +Y_{m_{4},m_{5}}(t_{3},t_{2}). \label{decoherence1}
\end{align}

In Eq.~\ref{SEresponse19}, the quantity $\big\langle \bar{D}_\text{SE}(t_{1},t_{2},t_{3}) \big\rangle$ is the decoherence term $\bar{D}_\text{SE}(t_{1},t_{2},t_{3})$ traced with respect to the bath degrees of freedom, given as
 \begin{align}
\big\langle \bar{D}_\text{SE}(t_{1},t_{2},t_{3}) \big\rangle & = \big\langle X^{\dag}_{m_{2}}(t_{1})+X^{\dag}_{m_{3}}(t_{2})+X_{m_{4}}(t_{3})\nonumber \\& +X_{m_{5}}(t_{2})+Y_{m_{2},m_{3}}(t_{1},t_{2}) \nonumber \\ & -Y_{m_{2},m_{4}}(t_{1},t_{3}) -Y_{m_{2},m_{5}}(t_{1},t_{2})\nonumber \\ & -Y_{m_{3},m_{4}}(t_{2},t_{3})-Z_{m_{3},m_{5}}(t_{2}) \nonumber   \\ & +Y_{m_{4},m_{5}}(t_{3},t_{2}) \big\rangle. \label{decoherence1a}
\end{align}

\section*{Appendix B: Population Relaxation in the Stimulated Emission Pathway} \label{A2}

$\bar{P}_\text{SE}(t_{1},t_{2},t_{3})$, 
when written out in full, contains several second 
order time-ordered terms. 
These involve integration, with respect to two 
different time arguments, $t$ and $s$, over an 
integrand, which, when traced over, has the general form
\begin{align}
\big\langle H_{na,kl}(t) H_{na,mn}(s) 
\big\rangle & \approx \sum_{j}A^{j}_{k,l}
(\textbf{0})A^{j}_{m,n}(\textbf{0}) \nonumber \\ 
& e^{i \triangle \varepsilon_{kl} t} 
e^{i \triangle \varepsilon_{mn} s} 
\big\langle \hat{P}_{j}(t)\hat{P}_{j}(s) 
\big\rangle. 
\label{poplnrelxn3}
\end{align}
Here, $H_{na}(t)$ is defined in the 
interaction picture, where $H_{0}$ 
is given in Eq.~\ref{H0}, 
$H_{na}$ in Eq.~\ref{Hna} and 
$\triangle \varepsilon_{kl}=\varepsilon_{k}-\varepsilon_{l}$. 
The approximation in Eq.~\ref{poplnrelxn3} 
arises because we use 
$e^{i H_{0} t}|m \rangle \approx 
e^{i \varepsilon_{m} t}|m \rangle$. 
$\big\langle P_{j}(t)P_{j}(s) \big\rangle$ 
can be easily evaluated to give
\begin{align}
\big\langle P_{j}(t)P_{j}(s) \big\rangle &=
\int^{\infty}_{0}d \omega S(\omega) \omega^{2} \nonumber\\
& \times \bigg(\frac{e^{-i \omega (t-s)}}{1-e^{-\beta \omega}}
+\frac{e^{i \omega (t-s)}}{e^{\beta \omega}-1}\bigg). 
\label{poplnrelxn4}
\end{align}
Therefore, we have
\begin{align}
&\big\langle H_{na,kl}(t) H_{na,mn}(s) 
\big\rangle \approx \sum_{j}A^{j}_{k,l}
(\textbf{0})A^{j}_{m,n}(\textbf{0})\nonumber \\ 
& \times \int^{\infty}_{0}d \omega S(\omega) 
\omega^{2}\bigg(\frac{e^{i(\triangle \varepsilon_{kl}- \omega)t}
e^{i(\triangle \varepsilon_{mn}+ \omega)s}}{1-e^{-\beta \omega}} \nonumber \\
&+\frac{e^{i(\triangle \varepsilon_{kl}+ \omega)t}
e^{i(\triangle \varepsilon_{mn}- \omega)s}}
{e^{\beta \omega}-1}\bigg). 
\label{poplnrelxn5}
\end{align}
The integration in Eq.~\ref{poplnrelxn5} is 
easily performed numerically. 
The population relaxation term for the SE pathway, 
$\bar{P}_\text{SE}(t_{1},t_{2},t_{3})$, again, contains four types of terms:\\
(a) a first order term, $\int^{t}_{0}dt'H_{{na},mn}(t')$, 
which, when traced over, gives zero as 
$\langle \hat{P}_{j}(t) \rangle=0$,\\
(b) a time-ordered second order term
\begin{equation}
L_{m,n}(t)=\int^{t}_{0}dt'\int^{t'}_{0}dt''
(H_{na}(t')H_{na}(t''))_{mn}, 
\label{L}
\end{equation}
(c) a second order term given by a product 
of two first order terms with different 
time arguments $t$ and $s$
\begin{align}
N_{k,l,m,n}(t,s)&=\int^{t}_{0}dt'\int^{s}_{0}dt''
(H_{{na},kl}(t') \nonumber \\ 
&\times H_{{na},mn}(t'')),
\label{N}
\end{align}
(d) a second order term given by a product of 
two first order terms with the same 
time argument $t$
\begin{align}
O_{k,l,m,n}(t)&=\int^{t}_{0}dt'\int^{t}_{0}dt''
(H_{{na},kl}(t') \nonumber \\ & \times H_{{na},mn}(t'')).
\label{O}
\end{align}

The second order terms are easily evaluated 
numerically, using Mathematica.
$\bar{P}_\text{SE}(t_{1},t_{2},t_{3})$ is given as,
\begin{align}
\bar{P}_\text{SE}(t_{1},t_{2},t_{3})&=
\delta_{m_{1},m_{2}}\delta_{m_{2},m_{3}}
\delta_{m_{4},m_{5}}L_{m_{5},m_{6}}(t_{2}) \nonumber \\
&+\delta_{m_{1},m_{2}}\delta_{m_{2},m_{3}}
\delta_{m_{5},m_{6}}L_{m_{4},m_{5}}(t_{3}) \nonumber \\ 
&+\delta_{m_{1},m_{2}}\delta_{m_{4},m_{5}}
\delta_{m_{5},m_{6}}L^{\dag}_{m_{2},m_{3}}(t_{2})\nonumber  \\
&+\delta_{m_{2},m_{3}}\delta_{m_{4},m_{5}}
\delta_{m_{5},m_{6}}L^{\dag}_{m_{1},m_{2}}(t_{1}) \nonumber \\ 
&+\delta_{m_{1},m_{2}}
\delta_{m_{2},m_{3}}N_{m_{4},m_{5},m_{5},m_{6}}(t_{3},t_{2})\nonumber \\ 
&-\delta_{m_{1},m_{2}}
\delta_{m_{4},m_{5}}O_{m_{2},m_{3},m_{5},m_{6}}(t_{2}) \nonumber \\
&-\delta_{m_{2},m_{3}}
\delta_{m_{4},m_{5}}N_{m_{1},m_{2},m_{5},m_{6}}(t_{1},t_{2})\nonumber \\
&-\delta_{m_{1},m_{2}}
\delta_{m_{5},m_{6}}N_{m_{2},m_{3},m_{4},m_{5}}(t_{2},t_{3})\nonumber \\
&-\delta_{m_{2},m_{3}}
\delta_{m_{5},m_{6}}N_{m_{1},m_{2},m_{4},m_{5}}(t_{1},t_{3})\nonumber \\
&+\delta_{m_{4},m_{5}}
\delta_{m_{5},m_{6}}N_{m_{1},m_{2},m_{2},m_{3}}(t_{1},t_{2}). 
\label{poplnrelxn1}
\end{align}
Solving Eq.~\ref{poplnrelxn1} is easy but can be 
expensive as the number of levels increases. 
However, it is worth noting that we have 
already incorporated memory/coherence effects 
in the expression for decoherence and 
Eq.~\ref{poplnrelxn1} contains only the 
incoherent population relaxation effects. 
Therefore, we make the approximation of 
neglecting the coupling of population 
relaxation effects during various time 
intervals (the $N$ terms) and use only 
the terms which contain population relaxation 
happening during one time interval 
(the $L$ and $O$ terms). 
Eq.~\ref{poplnrelxn1}, therefore, reduces to
\begin{align}
\bar{P}_\text{SE}(t_{1},t_{2},t_{3})& \approx 
\delta_{m_{1},m_{2}}\delta_{m_{2},m_{3}}
\delta_{m_{4},m_{5}}L_{m_{5},m_{6}}(t_{2}) \nonumber \\ 
& +\delta_{m_{1},m_{2}}\delta_{m_{2},m_{3}}
\delta_{m_{5},m_{6}}L_{m_{4},m_{5}}(t_{3}) \nonumber  \\ 
& +\delta_{m_{1},m_{2}}\delta_{m_{4},m_{5}}
\delta_{m_{5},m_{6}}L^{\dag}_{m_{2},m_{3}}(t_{2}) \nonumber \\
& +\delta_{m_{2},m_{3}}\delta_{m_{4},m_{5}}
\delta_{m_{5},m_{6}}L^{\dag}_{m_{1},m_{2}}(t_{1}) \nonumber \\
& -\delta_{m_{1},m_{2}}
\delta_{m_{4},m_{5}}O_{m_{2},m_{3},m_{5},m_{6}}(t_{2}). 
\label{poplnrelxn2}
\end{align}
In Eq.~\ref{SEresponse19}, 
the quantity 
$\big\langle \bar{P}_\text{SE}(t_{1},t_{2},t_{3}) \big\rangle$ 
is the population relaxation term 
$\bar{P}_\text{SE}(t_{1},t_{2},t_{3})$ traced with 
respect to the bath degrees of freedom, given as
\begin{align}
\big\langle \bar{P}_\text{SE}(t_{1},t_{2},t_{3}) \big\rangle & \approx 
\big\langle \delta_{m_{1},m_{2}}\delta_{m_{2},m_{3}}
\delta_{m_{4},m_{5}}L_{m_{5},m_{6}}(t_{2}) \nonumber \\ 
& +\delta_{m_{1},m_{2}}\delta_{m_{2},m_{3}}
\delta_{m_{5},m_{6}}L_{m_{4},m_{5}}(t_{3}) \nonumber  \\ 
& +\delta_{m_{1},m_{2}}\delta_{m_{4},m_{5}}
\delta_{m_{5},m_{6}}L^{\dag}_{m_{2},m_{3}}(t_{2}) \nonumber \\
& +\delta_{m_{2},m_{3}}\delta_{m_{4},m_{5}}
\delta_{m_{5},m_{6}}L^{\dag}_{m_{1},m_{2}}(t_{1}) \nonumber \\
& -\delta_{m_{1},m_{2}}
\delta_{m_{4},m_{5}}O_{m_{2},m_{3},m_{5},m_{6}}(t_{2})\big\rangle. 
\label{poplnrelxn2a}
\end{align}
Evaluating $\bar{P}_\text{SE}(t_{1},t_{2},t_{3})$, thus, 
requires at most $5$ numerical integrations 
(see Eq.~\ref{poplnrelxn2}).





\section*{Appendix C: Ground State Bleaching Response Function} \label{A3}
We provide a brief derivation of the GSB response function here. We have, from Eq.~\ref{GSBRF1},
\begin{align}
R_\text{GSB}(t_{3},t_{2},t_{1})&=\sum_{\{m\}}\mu_{gm_{1}} 
\mu_{m_{2}g} \mu_{gm_{3}} \mu_{m_{4}g} \nonumber \\ & e ^{i (\varepsilon_{m_{2}}-\epsilon_{g})t_{1}}e^{-i (\varepsilon_{m_{3}}-\epsilon_{g})t_{3}} \nonumber \\ & F_{\text{GSB}}(t_{1},t_{2},t_{3}), \label{GSBRF3a}
\end{align}
where
\begin{align}
F_{\text{GSB}}(t_{1},t_{2},t_{3})& = \big\langle (1-D_{\text{GSB}}(t_{1},t_{2},t_{3})) \nonumber \\ &(\delta_{m_{1},m_{2}}\delta_{m_{3}m_{4}}-P_{\text{GSB}}(t_{1},t_{2},t_{3})) \big\rangle, \label{GSBRF4}
\end{align}
where
\begin{align}
D_{\text{GSB}}(t_{1},t_{2},t_{3})&=1-(\hat{T}^{\dag}e^{i\int^{t_{1}}_{0}dt'\nabla_{\textbf{Q}} \varepsilon_{m_{2}}(\textbf{Q}(t'))\cdot \textbf{Q}(t')}) \nonumber \\ & \times (\hat{T}e^{-i\int^{t_{3}}_{0}dt'\nabla_{\textbf{Q}} \varepsilon_{m_{3}}(\textbf{Q}(t'))\cdot \textbf{Q}(t')}), \label{GSBRF6}
\end{align}
and
\begin{align}
P_{\text{GSB}}(t_{1},t_{2},t_{3})&=\delta_{m_{1},m_{2}}\delta_{m_{3}m_{4}} \nonumber   \\ & -(\hat{T}^{\dag}e^{i\int^{t_{1}}_{0}dt'H_{na}(t')})_{m_{1},m_{2}} \nonumber \\ &(\hat{T}e^{-i\int^{t_{3}}_{0}dt'H_{na}(t')})_{m_{3},m_{4}}. \label{GSBRF7}
\end{align}
Again, there could be two cases:

1. $\delta_{m_{1},m_{2}}\delta_{m_{3},m_{4}}=1$. We neglect the coupling between decoherence and population relaxation and then use the second order cumulant expansion to obtain
\begin{align}
F_{\text{GSB}}(t_{1},t_{2},t_{3}) &= \big\langle(1-D_{\text{GSB}}(t_{1},t_{2},t_{3})) \nonumber \\ & \times (1-P_{\text{GSB}}(t_{1},t_{2},t_{3}))\big\rangle \nonumber \\ & \approx e^{-\big(\big\langle \bar{D}_{\text{GSB}}(t_{1},t_{2},t_{3})\big\rangle+\big\langle \bar{P}_{\text{GSB}}(t_{1},t_{2},t_{3})\big\rangle \big)}. \label{GSBRF9}
\end{align}
$\bar{D}_{\text{GSB}}(t_{1},t_{2},t_{3})$ and $\bar{P}_{\text{GSB}}(t_{1},t_{2},t_{3})$ are defined below.

2. $\delta_{m_{1},m_{2}}\delta_{m_{3},m_{4}}=0$. In a way similar to the SE pathway, after decoupling decoherence and population relaxation and using the second order cumulant expansion, we have

\begin{equation}
F_{\text{GSB}}(t_{1},t_{2},t_{3}) =\big(1 -e^{\big\langle \bar{P}^{GSB}(t_{1},t_{2},t_{3})\big\rangle}\big). \label{GSBRF11}
\end{equation}

Here, 

\begin{align}
\bar{D}_{\text{GSB}}(t_{1},t_{2},t_{3})&=X^{\dag}_{m_{2}}(t_{1}) +X_{m_{3}}(t_{3}) \nonumber \\ &-Y_{m_{2},m_{3}}(t_{1},t_{3}), \label{GSBRF12}
\end{align}
where  $X_{m}(t)$ and $Y_{m,n}(t,s)$ have been defined in Eqs.~\ref{X}-\ref{Y} before.
Also,
\begin{align}
\bar{P}_{\text{GSB}}(t_{1},t_{2},t_{3})&=\delta_{m_{1},m_{2}}L_{m_{3},m_{4}}(t_{3}) \nonumber \\ & +\delta_{m_{3},m_{4}}L^{\dag}_{m_{1},m_{2}}(t_{1})\nonumber \\ &-N_{m_{1},m_{2},m_{3},m_{4}}(t_{1},t_{3}), \label{GSBRF13}
\end{align}
where $L_{m,n}(t)$ and $N_{k,l,m,n}(t,s)$ are defined in Eqs.~\ref{L}-\ref{N}. We could  neglect the coupling between population relaxations during different time intervals to obtain
\begin{align}
\bar{P}_{\text{GSB}}(t_{1},t_{2},t_{3}) & \approx \delta_{m_{1},m_{2}}L_{m_{3},m_{4}}(t_{3})\nonumber \\ &+\delta_{m_{3},m_{4}}L^{\dag}_{m_{1},m_{2}}(t_{1}). \label{GSBRF14}
\end{align}




\section*{Appendix D: Excited State Absorption Response Function} \label{A4}

We provide a brief derivation of the ESA response function here. We have from Eq.~\ref{ESARF3},
\begin{align}
R^{*}_\text{ESA}(t_{3},t_{2},t_{1})&= 
\sum_{\{m,n\}}\mu_{gm_{1}} \mu_{m_{4}n_{1}} 
\mu_{n_{2}m_{5}} \mu_{m_{6}g} \nonumber\\ & e^{i (\varepsilon_{m_{2}}-\epsilon_{g}) t_{1}}e^{i (\varepsilon_{m_{3}}-\varepsilon_{m_{5}}) t_{2}} \nonumber \\ & e^{i (\varepsilon_{m_{4}}-\varepsilon_{n_{1}}) t_{3}} F_{\text{ESA}}(t_{1},t_{2},t_{3}), \label{ESARF3a}
\end{align}
where
\begin{align}
F_{\text{ESA}}(t_{1},t_{2},t_{3})&=\big\langle (1-D_{\text{ESA}}(t_{1},t_{2},t_{3})) \nonumber \\ &\times (\delta_{m_{1},m_{2}}\delta_{m_{2},m_{3}}\delta_{m_{3}m_{4}}\delta_{n_{1},n_{2}}\delta_{m_{5},m_{6}}\nonumber \\ & -P_{\text{ESA}}(t_{1},t_{2},t_{3})) \big\rangle. \label{ESARF4}
\end{align}
Here,
\begin{align}
D_{\text{ESA}}(t_{1},t_{2},t_{3})&=1-(\hat{T}^{\dag}e^{i\int^{t_{1}}_{0}dt'\nabla_{\textbf{Q}} \varepsilon_{m_{2}}(\textbf{Q}(t'))\cdot \textbf{Q}(t')})\nonumber \\ & \times (\hat{T}^{\dag}e^{i\int^{t_{2}}_{0}dt'\nabla_{\textbf{Q}} \varepsilon_{m_{3}}(\textbf{Q}(t'))\cdot \textbf{Q}(t')}) \nonumber \\ & \times (\hat{T}^{\dag}e^{i\int^{t_{3}}_{0}dt'\nabla_{\textbf{Q}} \varepsilon_{m_{4}}(\textbf{Q}(t'))\cdot \textbf{Q}(t')}) \nonumber \\ & \times (\hat{T}e^{-i\int^{t_{3}}_{0}dt'\nabla_{\textbf{Q}} \varepsilon_{n_{1}}(\textbf{Q}(t'))\cdot \textbf{Q}(t')})\nonumber \\ & \times (\hat{T}e^{-i\int^{t_{2}}_{0}dt'\nabla_{\textbf{Q}} \varepsilon_{m_{5}}(\textbf{Q}(t'))\cdot \textbf{Q}(t')}), \label{ESARF6}
\end{align}
and
\begin{align}
P_{\text{ESA}}(t_{1},t_{2},t_{3})&=\delta_{m_{1},m_{2}}\delta_{m_{2},m_{3}}\delta_{m_{3}m_{4}}\delta_{n_{1},n_{2}}\delta_{m_{5},m_{6}} \nonumber \\ & -(\hat{T}^{\dag}e^{i\int^{t_{1}}_{0}dt'H_{na}(t')})_{m_{1},m_{2}}\nonumber \\ & \times (\hat{T}^{\dag}e^{i\int^{t_{2}}_{0}dt'H_{na}(t')})_{m_{2},m_{3}} \nonumber \\ &  \times (\hat{T}^{\dag}e^{i\int^{t_{3}}_{0}dt'H_{na}(t')})_{m_{3},m_{4}}\nonumber \\ &  \times (\hat{T}e^{-i\int^{t_{3}}_{0}dt'H_{na}(t')})_{n_{1},n_{2}}\nonumber \\ &  \times (\hat{T}e^{-i\int^{t_{2}}_{0}dt'H_{na}(t')})_{m_{5},m_{6}}. \label{ESARF7}
\end{align}




The Kronecker-delta constraints give us two cases, as before.

1. $\delta_{m_{1},m_{2}}\delta_{m_{2},m_{3}}\delta_{m_{3}m_{4}}\delta_{n_{1},n_{2}}\delta_{m_{5},m_{6}}=1$. This gives
\begin{align}
F_{\text{ESA}}(t_{1},t_{2},t_{3}))&= \big\langle(1-D_{\text{ESA}}(t_{1},t_{2},t_{3}))\nonumber \\ & \times (1-P_{\text{ESA}}(t_{1},t_{2},t_{3}))\big\rangle \nonumber \\ & \approx e^{-\big(\big\langle \bar{D}_{\text{ESA}}(t_{1},t_{2},t_{3})\big\rangle+\big\langle \bar{P}_{\text{ESA}}(t_{1},t_{2},t_{3})\big\rangle \big)}. \label{ESARF9}
\end{align}
$ \bar{D}_{\text{ESA}}(t_{1},t_{2},t_{3})$, $\bar{P}_{\text{ESA}}(t_{1},t_{2},t_{3})$ are defined below.

2. $\delta_{m_{1},m_{2}}\delta_{m_{2},m_{3}}\delta_{m_{3}m_{4}}\delta_{n_{1},n_{2}}\delta_{m_{5},m_{6}}=0$. Using second order cumulant expansion for decoherence and population relaxation and neglecting the coupling between decoherence and population relaxation, we have
\begin{align}
F_{\text{ESA}}(t_{1},t_{2},t_{3}))& \approx \big(1-e^{\big\langle \bar{P}_{\text{ESA}}(t_{1},t_{2},t_{3})\big\rangle}\big). \label{ESARF11}
\end{align}
Here,
\begin{align}
&\bar{D}_{\text{ESA}}(t_{1},t_{2},t_{3})=X^{\dag}_{m_{2}}(t_{1})+X^{\dag}_{m_{3}}(t_{2})+ \nonumber  \\ & X^{\dag}_{m_{4}}(t_{3})+X_{n_{1}}(t_{3})+X_{m_{5}}(t_{2})+Y_{m_{2},m_{3}}(t_{1},t_{2})\nonumber  \\ &+Y_{m_{2},m_{4}}(t_{1},t_{3})-Y_{m_{2},n_{1}}(t_{1},t_{3})-Y_{m_{2},m_{5}}(t_{1},t_{2})\nonumber  \\ &+Y_{m_{3},m_{4}}(t_{2},t_{3})-Y_{m_{3},n_{1}}(t_{2},t_{3})-Z_{m_{3},m_{5}}(t_{2}) \nonumber  \\ & -Z_{m_{4},n_{1}}(t_{3})-Y_{m_{4},m_{5}}(t_{3},t_{2})+Y_{n_{1},m_{5}}(t_{3},t_{2}), \label{ESARF12}
\end{align}
where $X_{m,n}(t)$, $Y_{m,n}(t,s)$ and $Z_{m,n}(t)$ are defined in Eqs.~\ref{X}-\ref{Z}. The second order population relaxation term, after neglecting the couplings during different time intervals, is given as
\begin{align}
\bar{P}_{\text{ESA}}(t_{1},t_{2},t_{3})&= \nonumber  \\ & \delta_{m_{2},m_{3}}\delta_{m_{3},m_{4}}\delta_{m_{5},m_{6}}\delta_{n_{1},n_{2}}L^{\dag}_{m_{1},m_{2}}(t_{1})\nonumber  \\ &+\delta_{m_{1},m_{2}}\delta_{m_{3},m_{4}}\delta_{m_{5},m_{6}}\delta_{n_{1},n_{2}}L^{\dag}_{m_{2},m_{3}}(t_{2})\nonumber  \\ &+\delta_{m_{1},m_{2}}\delta_{m_{2},m_{3}}\delta_{m_{5},m_{6}}\delta_{n_{1},n_{2}}L^{\dag}_{m_{3},m_{4}}(t_{3})\nonumber  \\ &+\delta_{m_{1},m_{2}}\delta_{m_{2},m_{3}}\delta_{m_{3},m_{4}}\delta_{n_{1},n_{2}}L_{m_{5},m_{6}}(t_{2})\nonumber  \\ &+\delta_{m_{1},m_{2}}\delta_{m_{2},m_{3}}\delta_{m_{3},m_{4}}\delta_{m_{5},m_{6}}L_{n_{1},n_{2}}(t_{3})\nonumber  \\ &-\delta_{m_{1},m_{2}}\delta_{m_{3},m_{4}}\delta_{n_{1},n_{2}}O_{m_{2},m_{3},m_{5},m_{6}}(t_{2})\nonumber  \\ &-\delta_{m_{1},m_{2}}\delta_{m_{2},m_{3}}\delta_{m_{5},m_{6}}O_{m_{3},m_{4},n_{1},n_{2}}(t_{3}), \label{ESARF13}
\end{align}
where $L_{m,n}(t)$ and $O_{k,l,m,n}(t)$ are defined in Eqns. (\ref{L})-(\ref{O}).





                            
\end{document}